# Structural Performance of Laterite soil Stabilised with Cement and Blue Gum (Eucalyptus Globulus) Wood Ash for Use as a Road base Material

Blaise Dabou[1], Christopher Kanali[2], Zachary Abiero-Gariy[3]

[1]Pan African University Institute for Basic Sciences, Technology and Innovation hosted at Jomo Kenyatta University of Agriculture and Technology, Kenya

[2]Department of Agricultural and Biosystems Engineering, Jomo Kenyatta University of Agriculture and Technology, Kenya

[3]Department of Civil Engineering, Jomo Kenyatta University of Agriculture and Technology, Kenya

[1]dabou.blaise@students.jkuat.ac.ke, [2]ckanali@jkuat.ac.ke, [3]zagariy@yahoo.co.uk

**Abstract** — This study examines the effect of partially replacing cement with blue gumwood ash (BGWA) in stabilizing laterite soil to be used as a potential road base material. Initially, Ordinary Portland Cement was introduced to the soil at varying contents from 0 to 12% in steps of 3% by weight of the soil sample. Each wet sample of the soil was subjected to the Californian bearing ratio (CBR) and unconfined compressive strength (UCS) tests in determining the optimal soil-cement mix. Results show that CBR and UCS values increased as cement content increased, and a 6% cement content corresponding to a UCS value of 2.88 MPa at 7 days of curing met the specifications of the Overseas Road Note 31 to be used in the construction of road bases. The second treatment involved partially replacing the 6% cement content with BGWA in decreasing steps of 1%. Peak CBR value of 348% at 2% BGWA content (>160% recommended by Kenya road design manual) and UCS value of 2.99 MPa at 7 days of curing were obtained. Thus, BGWA can partially replace cement in stabilizing laterite soil for use in the construction of road bases as per the Overseas Road Note 31 specifications.

## I. INTRODUCTION

The high cost of road construction and maintenance has a negative impact on the economic development of the vast majority of developing countries, the majority of which are in Africa, such as Mali and Kenya. Mali has a poor road network as a result of the high incidences and frequency of road pavement failures caused by poor quality soil material used. Studies show that only 40% of the road network was in good condition in Mali in 2011, according to the national roads directorate[1]. The solution to this problem is to develop alternative ways of improving the quality of locally available in-situ construction material to reduce the cost of transportation. Such improvement can be achieved through soil stabilization which consists of mixing one or more materials with the in-situ material then compacting it to attain high strength and durability. The process of soil stabilization using cement and lime has long been used to improve the properties of clayey soils. These stabilizers have become very expensive, and their negative impact on the environment has reduced their usage. The primary goal of this research is to determine the feasibility of using blue gumwood ash (BGWA), a by-product from tea factories, bakeries, and households, for stabilization of laterite soil for the construction of road base. The BGWA is used to partially replace cement during the laterite soil stabilization process.

Wood has long been used the world over for energy purposes. Until now, the vast majority of African countries depend on wood to supply their basic energy demands, resulting in the generation of ash. Wood fuel consumption was estimated at 60 to 80% of the primary energy[2]. Wood ash is the powdery residue that remains after burning wood in a household fireplace or in a factory (such as tea factories in Kenya). It is normally deposited as waste and can litter the environment impacting the plant's aesthetic conditions. According to [3], cement production is one of the main sources of environmental pollution worldwide due to the emission of CO2. Almost one ton of CO2 is released to the environment for every ton of cement produced [4]. Hence, using wood ash as a soil additive to partially replace cement will minimize cement used in the soil stabilization process and decrease the environmental pollution.

"Ref. [5]" investigated the effect of wood ash on laterite soil stabilization. The soil was stabilized with different proportions of wood ash from 0 to 10% in steps of 2%. Atterberg limits, Proctor compaction, and CBR tests were performed on unstabilized and stabilized soil samples. The addition of wood ash increased the optimum moisture content from 7.7 % at 0% wood ash content to 17.27 % at 10% wood ash content, resulting in a decrease in maximum dry density from 1970 kg/m3 to 1840 kg/m3. The value of the CBR increased from 31.44% to 46.72% at 0% and 10% wood ash content, respectively.

In "Ref. [6]", ash derived from banana leaves was used to stabilize a lateritic soil from Akure, Nigeria. The ash was introduced to the soil in varying proportions ranging from 2% to 10% in steps of 2% by weight of the soil sample. Proctor compaction, CBR, and UCS tests were performed on both unstabilized and stabilized soils. According to the results of the strength tests, ash from banana leaves improved the soil's strength parameters. As a result, the author concluded that ash from banana leaves could be used to economically improve lateritic soil for subgrade purposes.

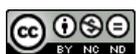

"Ref. [7]" investigated the effect of oil palm fonds (OPFA) ashes on lateritic soil. The soil was stabilized with varying proportions of oil palm fonds ash ranging from 2 to 10% in steps of 2% of the weight of the soil sample. The obtained results were then compared to the obtained results from the same soil stabilized with cement to determine the performance of oil palm fonds ashes in relation to cement. Results from the compaction test show the maximum dry densities were 2.02 kg/m$^3$ at 4% ashes of oil palm fonds content and 2.40 kg/m$^3$ at 6% cement content. The highest CBR values (32.6% and 87.32%) were found at 4% OPFA and 6% cement contents, which corresponded to the highest UCS values (236.86 kg/m$^3$ and 588.32 kg/m$^3$). The highest oxides in the oil palm fond ash, according to chemical analysis, were CaO (60.83%) and $SiO_2$ (33.67%).

The effect of ash from the ground-nut husk on the geotechnical properties of a lateritic soil was studied in [8]. Proctor compaction, CBR, and UCS tests were performed on both unstabilized and stabilized soil with ground-nut husk ash. According to the results of the tests, the CBR and UCS were significantly improved, but the maximum dry density did not improve because it decreased from 1960 kg/m3 to 1760 kg/m3 while the optimum moisture content increased from 12.70% to 14.95% all at 10% ground-nut husk ash content. It was therefore concluded that ground-nut husk ash stabilizes lateritic soil for subgrade and subbase purposes at low costs.

The effect of ash from rice husk ash (RHA) on the geotechnical properties of a laterite soil collected from Mainkunkele area of Minna, Nigeria, was investigated in [9]. The soil was stabilized with varying proportions of rice husk ash (2-8%) by the weight of dry soil. The effect of RHA on stabilized soil strength properties (i.e., CBR and UCS) was investigated. According to the findings, increasing the RHA content from a specific cement content resulted in a decrease in maximum dry density and an increase in optimum moisture content. CBR and UCS improved significantly as RHA content increased, with peak values observed between 4 and 6% RHA content. UCS was also found to increase with curing age. The author concluded that an RHA content of 4 to 6% admixed with a low cement content could effectively stabilize a laterite soil.

"Ref. [10]" investigated the effect of asphalt emulsion-cement mixture on three different lateritic samples A, B and C collected in their natural states. In order to determine the best mix ratio, asphalt emulsion was introduced into the soil sample in various contents ranging from 2 to 8% in steps of 2% of the weight of the soil sample, while the cement was varied over each percentage of asphalt emulsion at 0.5, 1 and 2 % of the weight of the soil sample. CBR and UCS tests were conducted on both untreated and treated soil samples using an asphalt emulsion-cement mixture. Results show that the addition of asphalt emulsion increased the soil sample's strengths. Positive results were further obtained with the addition of cement to the asphalt emulsion for stabilizing all soil samples. It is was observed that highly plastic soils did not respond favorably with stabilizing with only asphalt emulsion but was improved significantly with the stabilization using asphalt emulsion-cement mixture. The author concludes that asphalt emulsion-cement mixture improves well high plastic soils and leads to the reduction in the cost of stabilized soils, especially clayey soils, which require a high amount of cement for their treatment.

"Ref. [11]" examined the effect of bitumen emulsion mixed with cement on the geotechnical properties of a lateritic soil obtained from borrow pits in Kwali area, Nigeria. Three different percentages for mixing the bitumen emulsion with the cement were considered: 4%, 6%, and 8% of the weight of the soil sample. The bitumen emulsion was mixed with the cement in various proportions to form five additives. On both the stabilized and unstabilized soil samples, CBR and UCS tests were performed. CBR and UCS for soil samples A and B were 19.6% and 0.46 MPa, 22.6%, and 0.95 MPa, respectively. A significant improvement in the soil CBR was observed for sample A at 4% additives of the mix proportions used. The corresponding UCS (i.e., for 4% additives) was also improved since it increased from 0.64 MPa to 1.33 MPa. While CBR for sample B increased from 78.4% to 288.1% at 8% additives, and the corresponding UCS from 0.48 MPa to 2.45 MPa. It was observed both CBR and UCS increased with an increase in cement content. The author concluded stabilization using bitumen emulsion-cement mixture improves soil strength.

Based on the above previous works on the potential use of agricultural and industrial wastes referred to as non-traditional stabilizers, including wood ash, it is possible to conclude that these by-products can be used as stabilizers mainly for laterite and lateritic soils. They improve the strength parameters of these soil types, especially ashes, due to the pozzolanic reaction occurring between calcium hydroxide containing in the soil and the ash, thus forming cementitious materials. In most cases, they fail to improve the maximum dry density of the soil and the strength parameters (CBR and UCS) when the ash content increases over the soil or the binder used; this could be due to the lower density of these by-products in comparison to the density of the soil or binder (cement or lime). They fill soil voids and decrease the number of clay particles. They also require more water during their modification process to achieve desired properties which are not favorable, especially when the construction site is in a remote area where access to water is not at all easy.

## II. MATERIALS AND METHODS

### A. *Materials acquisition and Preparation*

This study's materials included laterite soil, Ordinary Portland cement (OPC), BGWA, and water. The laterite soil was procured locally around Jomo Kenyatta University of Agriculture and Technology (JKUAT), Kenya. The soil was taken from different points in a pile of laterite, indicating that it had already been disturbed. The soil was transported to the civil engineering laboratory of JKUAT and was then air-dried before use. The OPC was of nominal strength 42.5 MPa (CEMI) and was also purchased locally. Similarly, the BGWA was sourced locally from a landfill of Juja Pulp and Paper Ltd factory. On visual inspection, the ash appeared to be grey and came in powder form. The ash was sieved using a 0.3 mm BS sieve to remove lumps and was then kept in a bag ready for use.

### B. Data Collection Procedure

To begin with, the engineering properties of the laterite soil, OPC, and BGWA were determined based on the BS1377 standard[12]. Thereafter, the soil was first treated with cement only by adding to it varying proportions of cement from 0 to 12% in steps of 3% by weight of the soil. Previous studies adopted the same proportion for soil stabilization for possible use in roads construction [13]–[15]. Various tests were performed on the soil-cement samples in order to obtain the optimum cement content (x%) to be used in later treatment. The tests conducted included Atterberg limits, Proctor compaction, CBR, and UCS. All these tests were carried out as per BS1924[16].

The Proctor compaction test, which determines the relationship between MDD and OMC, was conducted at the modified energy level. Each soil-cement sample was thoroughly mixed using a trowel after adding 6% of water by weight of the mix of the air-dried soil sample and cement to achieve the required moisture content. The sample was then compacted in five (5) layers in a standard Proctor mold with an internal diameter of 100 mm and an internal height of 115 mm. Each layer received 27 blows/rams from the compaction hammer.

The CBR tests were performed on samples compacted at their OMCs obtained from the Proctor compaction test, which were then cured in a curing cabinet for 7 days before being soaked in water for another 7 days in accordance with BS1924[16]. On the other hand, the UCS tests were carried out on samples compacted at their OMCs in the standard Proctor mold. The samples were cured for 7 and 14 days under controlled conditions (in a curing cabinet with a relative humidity condition of 100%). They were then subjected to a uniaxial compression test at a rate of 0.2 m/s using a compression machine. At failure, the maximum load was recorded and the maximum compressive strength, which is the UCS value computed. Since the length to diameter ratio of the samples was less than 2, ASTM C39/C39M[17] correction factors which are found applicable to soil-cement according to the Portland Cement Association[18], were used to obtain samples real UCS values.

The second treatment involved partially replacing cement with BGWA in the soil-cement mixtures from the optimal content of x% in decreasing steps of 1%, of weight the soil (II. TABLE I) to form soil-cement-BGWA samples. The The soil-cement-BGWA samples were thoroughly mixed using a trowel after adding 6% of water by weight of the soil. These samples were subjected to the same tests as in the first treatment in accordance with BS1924[16]

Finally, the soil-cement and soil-cement-BGWA samples were subjected to durability tests. Durability was expressed in terms of the resistance of the samples to strength loss. This was achieved by dividing the UCS value obtained from stabilized samples cured for 7 days and soaked in water for another 7 days to that of the UCS value obtained from another set of stabilized samples cured for 14 days under controlled conditions as per BS1924[16]. This testing method was preferred to the wet-dry and freeze-thaw tests specified in the ASTM standard because it better represents field conditions in the study area. The same testing method was adopted in [19] to assess the durability of lateritic soil stabilized with eggshell and cement and in [20] to assess the durability of laterite soil stabilized with cement for use as a flexible pavement construction material.

Three samples were used each for UCS, durability, and specific gravity tests, while two were used for the CBR test. One sample was used for each of the other tests (i.e., grain size distribution, Atterberg limits, and the chemical analysis).

### II. TABLE I
MIX DESIGN USED FOR LATERITE SOIL-CEMEN-BGWA TREATMENT

| Proportions of mix design (%) | | |
|---|---|---|
| Laterite soil | Cement | Bluegum wood ash |
| 100 | 0 | 0 |
| 100-x | x | 0 |
| 100-x | x-1 | 1 |
| 100-x | x-2 | 2 |
| 100-x | x-3 | 3 |
| - | - | - |
| -. | - | - |
| 100-x | 0 | x |

### III. RESULTS AND DISCUSSIONS

#### A. Characterization of Laterite Soil and Bluegum Wood ash

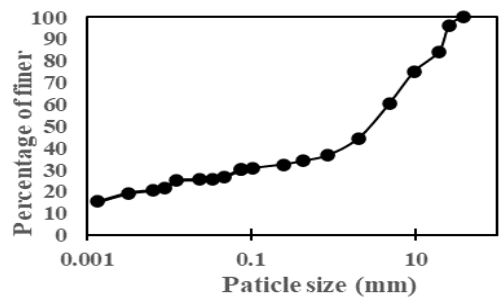

**Fig. 1 Grain size distribution curve of the soil**

III.TABLE summarizes the properties of laterite soil prior to the addition of any stabilizer, while Figure 1 shows the grain size distribution curve of the soil. According to the American Association of State Highway Transportation Officials (AASHTO) classification system, the overall engineering properties of the soil were classified as A-2-7. This reveals that the soil is unsuitable to be used as a road base construction material, necessitating the need to stabilize it[21], [22].

III.TABLE II and III. TABLE III shows the chemical compositions of laterite soil, cement, and BGWA. The main oxides found in the laterite soil were iron oxides, silica oxides, and aluminum oxides, which accounted for 50.13%, 30.81%, and 6.66%, respectively. The silica-to-sesquioxides ratio (Al2O3+SiO2) was adopted as an indication of the degree of laterization. This ratio was found to be 0.54, which is less than 1.33, implying that the soil is laterite [23]. The sum of the silica and sesquioxides contents in the BGWA is less than 70%, the minimum required by ASTM C 618-05[24], so it is not classified as pozzolana. The loss on ignition value for BGWA (12.24%) was slightly greater than 12%, the minimum required by ASTM C618-05 [24]. This could be attributed to the ash containing a significant amount of unburned carbon, which reduces its pozzolanic activity as a result of the uncontrolled heating process.

**III.TABLE I**
**ENGINEERING PROPERTIES OF THE LATERITE SOIL**

| Properties | Proportion/Value |
|---|---|
| Natural moisture content | 12.17±0.10% |
| Specific gravity | 2.67±0.01 |
| %Passing through BS sieve 75μ | 30.04% |
| Liquid limit | 41.74% |
| Plastic limit | 22.73% |
| Plasticity Index | 19.02% |
| AASHTO classification | A-2-7 |
| Californian bearing ratio (4 days soak) | 36% |
| Unconfined Compressive Strength | 0.58 MPa |

**III. TABLE II**
**CHEMICAL COMPOSITION OF THE SOIL**

| Constituents | Formula | Composition (%) |
|---|---|---|
| Iron | $Fe_2O_3$ | 50.13 |
| Silica | $SiO_2$ | 30.81 |
| Aluminium | $Al_2O_3$ | 6.66 |
| Manganese | MnO | 4.36 |
| Titanium | TiO | 2.98 |
| Calcium | CaO | 1.36 |
| Phosphorus | P2O | 1.07 |
| Potassium | K2O | 0.94 |
| Sulphur | S | 0.68 |
| Barium | Ba | 0.52 |
| Zirconium | Zr | 0.25 |

**III. TABLE III**
**CHEMICAL COMPOSITION OF OPC&BGWA**

| Oxides | Oxides contained in OPC (%) | Oxides contained in BGWA (%) |
|---|---|---|
| Calcium (CaO) | 82.71 | 80.77 |
| Potassium ($K_2O$) | 0.29 | 7.69 |
| Silica ($SiO_2$) | 6.09 | - |
| Phosphorus ($P_2O_5$) | 1.72 | 3.96 |
| Iron (Fe2O3) | 3.15 | 0.70 |
| Sulphur (S) | 3.11 | 1.15 |
| Aluminium ($Al_2O_3$) | 2.29 | 0.98 |
| Manganese (MnO) | - | 2.11 |
| Chlorine (Cl) | - | 1.91 |
| Strontium (Sr) | - | 0.36 |
| Titanium (Ti) | 0.27 | 0.11 |
| Zirconium (Zr) | - | 0.04 |
| Loss on Ignition (LOI) | 3.79 | 12.24 |

*B. Stabilization of Laterite Soil with Cement*

Results in III. TABLE IV and Figure 2 show that as cement content increased, the plasticity index decreased, indicating that the soil plasticity improved. The higher the plasticity index, the more clayey the soil is[25]. On the other hand, the results in III. TABLE V and Figure 3 show that the MDD and OMC did not change much as the cement content increased. This is due to the addition of small amounts of cement and the fact that the hydration phenomenon did not take place in a short period of time [26]. "Ref. [27]" reported a similar finding for cement stabilized soil. It was also observed (III.TABLE VI, Figure 4 and Figure 5) that the strength parameters (CBR and UCS) increased as the cement content and curing period increased (UCS). This is as a result of compaction, hydration of cement, and the pozzolanic reaction between cement and soil particles resulting in the formation of calcium silicate hydrate and calcium aluminate hydrate, both of which contribute to the formation of a compound that binds the soil particles over time[28].

The UCS at 7 days of curing is the most important strength criterion used for cement stabilized materials for road purposes. "Ref.[29]" stated that the UCS test should be used to determine the strength of cement stabilized soil base. Since the UCS values at 7 days of curing increased with cement content and that all the soil-cement mixes from 6% cement content satisfy the specification (1.5-3 MPa) of the Oversea Road Note 31[22] to be used in the construction of road bases, the 6% cement content was selected as the optimum cement content.

### III.TABLE IV
**ATTERBERG LIMITS AT VARYING CEMENT CONTENT**

| Mix design (%) | | Liquid limit (%) | Plastic limit (%) | Plasticity index (%) |
|---|---|---|---|---|
| Laterite soil | Cement | | | |
| 100 | 0 | 41.74 | 22.73 | 19.01 |
| 97 | 3 | 40.81 | 22.67 | 18.14 |
| 94 | 6 | 40.02 | 26.58 | 13.44 |
| 91 | 9 | 39.94 | 27.23 | 12.71 |
| 88 | 12 | 37.89 | 25.30 | 12.59 |

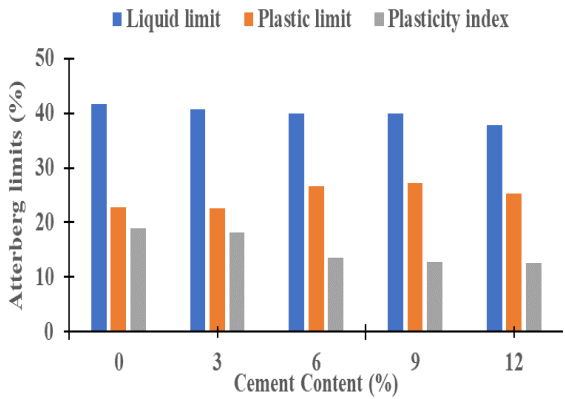

**Fig. 2 Variation of Atterberg limits with cement content**

### III.TABLE V
**MDD AND OMC AT VARYING CEMENT CONTENT**

| Mix design (%) | | MDD (g/cm³) | OMC (%) |
|---|---|---|---|
| Laterite soil | Cement | | |
| 100 | 0 | 1.97 | 12.54 |
| 97 | 3 | 1.97 | 13.06 |
| 94 | 6 | 1.96 | 11.67 |
| 91 | 9 | 1.97 | 11.55 |
| 88 | 12 | 1.98 | 12.49 |
| | Mean | 1.97±0.007 | 12.26±0.637 |

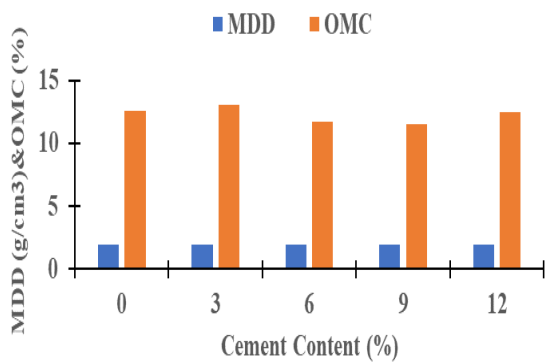

**Fig. 3 Variation of MDD and OMC with a variation of cement content**

### III.TABLE VI
**UCS AND CBR AT VARYING CEMENT CONTENT**

| Mix design (%) | | Soaked CBR (%) | UCS at 7 days (MPa) | UCS at 14 days (MPa) |
|---|---|---|---|---|
| Laterite soil | Cement | | | |
| 100 | 0 | 36 | 0.58 | 0.58 |
| 97 | 3 | 105 | 1.42 | 1.49 |
| 94 | 6 | 122 | 2.88 | 2.98 |
| 91 | 9 | 231 | 3.40 | 3.46 |
| 88 | 12 | 433 | 3.59 | 3.77 |

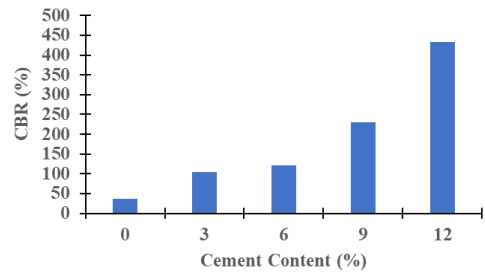

**Fig. 4 Variation of CBR with various cement contents**

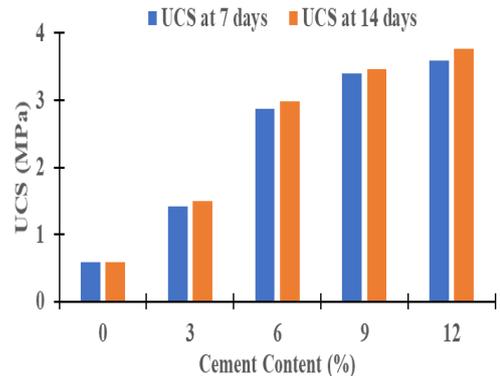

**Fig. 5 Variation of UCS with various soil-cement ratios**

### C. Stabilization of laterite soil with Cement and Blue Gum Wood Ash

During this treatment, the soil was stabilized with cement and with BGWA as a replacement of cement from the optimum cement content of 6% in decreasing steps of 1%. The results are presented in III. TABLE VII to XI. The results in III. TABLE VII shows that the value of the plasticity index decreased overall (up to 12.49%) with an increase in BGWA content (up to 2%) accompanied by a decrease in cement content from 6-4%. Thereafter, the index increased (up to 14.89%) with an increase in BGWA content (up to 4%) and with a further decrease in cement content up to 2%. The decrease in the plasticity index indicates that the cement-BGWA admixture improved the soil's plasticity. According to [25], the higher the plasticity index, the more clayey the soil is.

An increase in BGWA content from 0-6% alongside a decrease in cement content from 6-0% did not have much influence on the MDD (1.94±0.022 g/m3), but it increased the OMC from 11.67 to 14.4% (III.TABLE VIII and Figure 6). This may be attributed to the addition of small amounts of BGWA in the soil-cement mixture. An increase in OMC is attributed to the hydration phenomenon and the finesses of the stabilizers resulting in a demand for more water[30].

### III.TABLE VII
#### ATTERBERG LIMITS FOR DIFFERENT DESIGN MIXES

| Mix design (%) | | | Liquid limit (%) | Plastic limit (%) | Plasticity index (%) |
|---|---|---|---|---|---|
| Laterite soil | Cement | BGWA | | | |
| 100 | 0 | 0 | 41.74 | 22.73 | 19.01 |
| 94 | 6 | 0 | 40.02 | 26.58 | 13.44 |
| 94 | 5 | 1 | 40.47 | 27.5 | 12.97 |
| 94 | 4 | 2 | 39.27 | 26.78 | 12.49 |
| 94 | 3 | 3 | 41.01 | 26.77 | 14.24 |
| 94 | 2 | 4 | 41.24 | 26.35 | 14.89 |
| 94 | 1 | 5 | 40.40 | 26.70 | 13.70 |
| 94 | 0 | 6 | 40.51 | 26.4 | 14.11 |

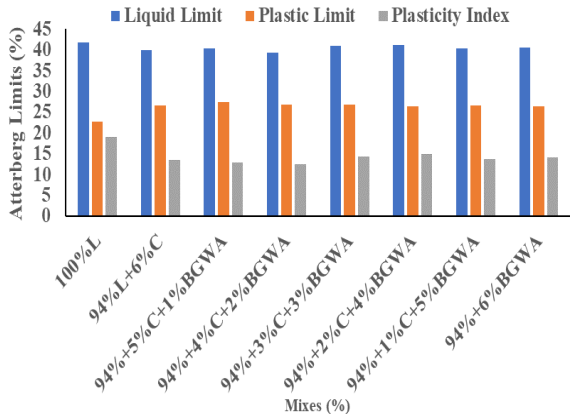

**Fig. 6 Variation of Atterberg limits for different design mixes of soil-cement-blue gumwood ash**

### III.TABLE VIII
#### MDD AND OMC FOR DIFFERENT DESIGN MIXES

| Mix design (%) | | | MDD (g/cm³) | OMC (%) |
|---|---|---|---|---|
| Laterite soil | Cement | BGWA | | |
| 100 | 0 | 0 | 1.97 | 12.54 |
| 94 | 6 | 0 | 1.96 | 11.67 |
| 94 | 5 | 1 | 1.96 | 13.50 |
| 94 | 4 | 2 | 1.92 | 14.00 |
| 94 | 3 | 3 | 1.92 | 14.00 |
| 94 | 2 | 4 | 1.92 | 13.60 |
| 94 | 1 | 5 | 1.94 | 14.40 |
| 94 | 0 | 6 | 1.92 | 14.40 |
| | | Mean | 1.94±0.02 | 13.51±0.96 |

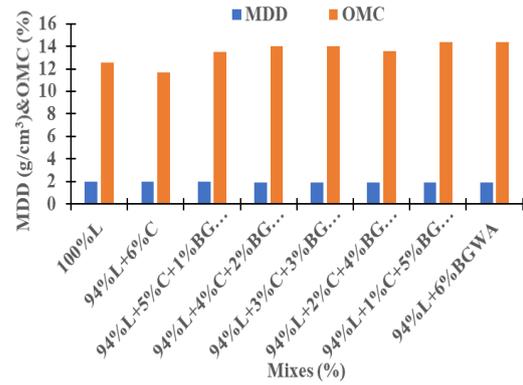

**Fig. 7 Variation of MDD and OMC for different design mixes of soil-cement-blue gum wood ash**

### III.TABLE IX
#### SOAKED CBR FOR DIFFERENT DESIGN MIXES

| Mix design (%) | | | CBR (%) |
|---|---|---|---|
| Laterite soil | Cement | BGWA | |
| 100 | 0 | 0 | 36 |
| 94 | 6 | 0 | 122 |
| 94 | 5 | 1 | 299 |
| 94 | 4 | 2 | 348 |
| 94 | 3 | 3 | 242 |
| 94 | 2 | 4 | 187 |
| 94 | 1 | 5 | 180 |
| 94 | 0 | 6 | 61 |

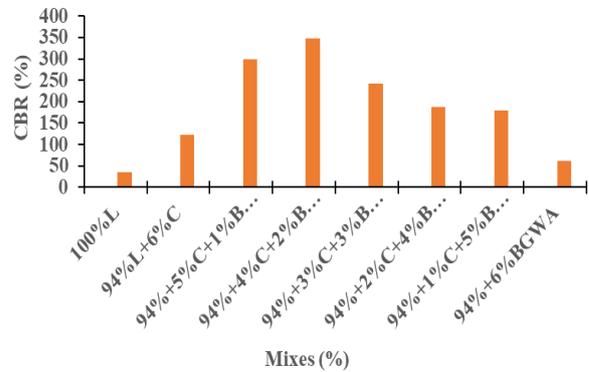

**Fig. 8 Variation of CBR values for different design mixes**

### III.TABLE X
#### UCS VALUES FOR DIFFERENT DESIGN MIXES AND CURING PERIODS

| Mix design (%) | | | UCS at 7 days (MPa) | UCS at 14 days (MPa) |
|---|---|---|---|---|
| Laterite soil | Cement | BGWA | | |
| 100 | 0 | 0 | 0.58 | 0.58 |
| 94 | 6 | 0 | 2.88 | 2.98 |
| 94 | 5 | 1 | 2.99 | 3.27 |

| 94 | 4 | 2 | 2.51 | 2.63 |
| 94 | 3 | 3 | 2.66 | 2.73 |
| 94 | 2 | 4 | 1.97 | 2.18 |
| 94 | 1 | 5 | 1.70 | 1.75 |
| 94 | 0 | 6 | 1.19 | 1.22 |

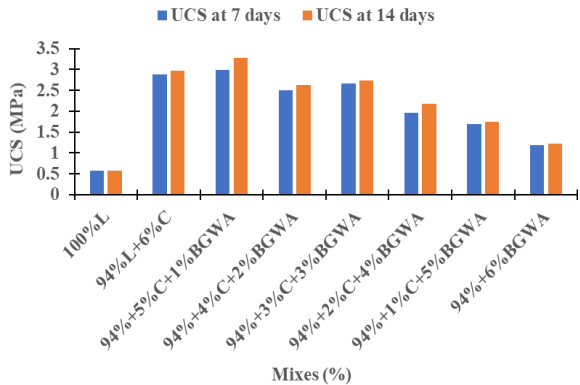

**Fig. 9 Variation of UCS values for different design mixes of soil-cement-blue gum wood ash and curing periods**

The results further show that soaked CBR increased to an optimum value of 348% at 4% cement content and 2% BGWA content (III.TABLE VIII). The chemical reaction between the stabilizers and the soil, complimented by compaction, may be responsible for the increase of the CBR. The CBR values obtained for 6-0% cement content and 0-5% BGWA content satisfy the specification (minimum 160%) of the Kenya road design manual to be used in the construction of road bases[21].

The UCS is an important soil testing method used to assess the quality of stabilized materials. "Ref. [29]" stated that the UCS test should be used to determine the strength of cement stabilized soil base.

The results in III. TABLE X and Figure 9 show that UCS values decreased from 2.88 to 1.19 MPa and from 2.98 to 1.22 MPa for 7- and 14-days curing, respectively, with an increase in BGWA content from 0-6% and a decrease in cement content from 6-0%.

The lower density (specific gravity) of the BGWA replacing those of the soil and cement may contribute to the decrease of the UCS. The values UCS at 7 days curing for 1-5% BGWA content and 5-1% cement content are within the range of 1.5-3 MPa specified in the Overseas Road Note 31[22].

The durability of the optimum soil-cement-BGWA admixtures under simulated tropical conditions was determined from the resistance of samples to strength loss. The results are shown in III. TABL XI and Figure 10. The durability was constant at a mean value of 81.94±1.974% for BGWA content from 0 to 3%; thereafter, it decreases significantly to 32.94% at 6% BGWA content. The durability for BGWA content between 0 and 2% was found to be satisfactory as values were above recommended minimum of 80%[20].

**III.TABLE XI**
**DURABILITY FOR DIFFERENT DESIGN MIXES**

| Mix Design (%) | | | Durability (%) |
|---|---|---|---|
| Laterite soil | Cement | BGWA | |
| 100 | 0 | 0 | - |
| 94 | 6 | 0 | 84.56 |
| 94 | 5 | 1 | 81.01 |
| 94 | 4 | 2 | 82.22 |
| 94 | 3 | 3 | 79.97 |
| 94 | 2 | 4 | 74.29 |
| 94 | 1 | 5 | 62.13 |
| 94 | 0 | 6 | 32.94 |

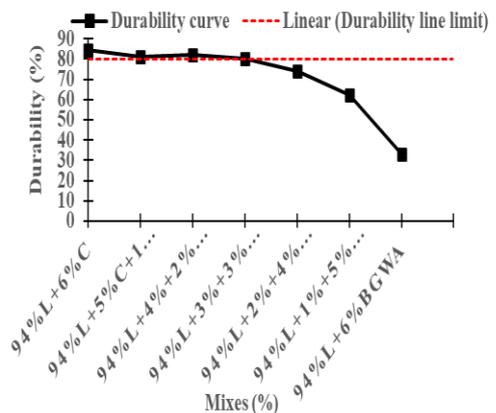

**Fig. 10 Variation of Durability for different design mixes**

### IV. CONCLUSIONS

The laterite soil used in this study contained 55.66% of gravel, 12.48% of sand, 4.93% of silt, and 26.93% of clay with a specific gravity of 2.67 and was classified A-2-7 based on the American Association of State Highways Transportation Officials (AASHTO) classification system. Atterberg limits of the laterite soil show the soil is clayey (Plasticity Index = 19.01%). On the other hand, Portland cement of nominal strength of 42.5 MPa with a specific gravity of 3.04 was used. The blue gumwood ash had a specific gravity of 2.47 and could not be classified as a pozzolana according to ASTM standards.

A 6% cement content in the soil-cement-water mixture was selected as the optimum content based on the unconfined compressive strength (UCS) values for 7 days of curing and based on the specification (1.5 to 3 MPa) of the Road & Development (1993) to be used in the construction of road bases. An increase in blue gum wood ash content from 0-6% alongside a decrease in cement content from 6-0% did not have much influence on the maximum dry density (1.94±0.022 g/m3) of the samples, but it increased the optimum moisture content from 11.67 to 14.4%. It was observed that the soaked Californian Bearing ratio (CBR) increased to an optimum value of 348% at 4% cement content and 2% blue gumwood ash. The UCS values decreased from 2.88 MPa to 1.19 MPa and from 2.98 MPa to 1.22 MPa for 7- and 14-days curing, respectively, with an increase in blue gumwood ash content from 0-6% and a decrease in cement content from 6-0%. The durability for blue gumwood ash content between 0 and 2% was found to be satisfactory as the values were above the recommended minimum of 80%.

Thus, it can be concluded that BGWA is an ideal material to partially replace cement in the process of laterite soil stabilization for road construction purposes at a cheap cost.

### ACKNOWLEDGMENT

The authors acknowledge the Pan African University Institute for Basic Sciences, Technology and Innovation and Jomo Kenyatta University of Agriculture and Technology for providing facilities and laboratory assistance throughout the completion of this research work. They also gratefully acknowledge African Union for their fund's support.

## REFERENCES

[1] BOAD, Etude sur l'entretien routier dans les pays de l'union économique et monétaire ouest africaine (UEMOA) : Bilan des 50 dernières années et perspectives., (2015). [Online]. Available: https://www.boad.org/wpcontent/uploads/2016/11/150813_agecet_diwi_rapport_prov_er_01.pdf.

[2] S. Amous, The Role of Wood Energy in Africa, (1999). http://www.fao.org/3/x2740e/x2740e00.htm#TopOfPage (accessed Mar. 14, 2021).

[3] A. Ekinci, M. Hanafi, and E. Aydin, Strength, stiffness, and microstructure of wood-ash stabilized marine clay, Minerals, 10(9) (2020) 1–23. doi: 10.3390/min10090796.

[4] M. Ayeldeen and M. Kitazume, Geotextiles and Geomembranes Using fiber and liquid polymer to improve the behavior of cement-stabilized soft clay, Geotext. Geomembranes, 45(6) (2017) 592–602, 2017, doi: 10.1016/j.geotexmem.2017.05.005.

[5] R. Abdulwahab, B. Ibitoye, and M. Akinleye, The Effects of Wood ash on the Geotechnical Properties of Lateritic Soil, J. Res. Inf. Civ. Eng.,October, 15 (2018)

[6] E. S. Nnochiri and O. Aderinlewo, Geotechnical Properties of Lateritic Soil Stabilized with Banana Leaves Ash, FUOYE J. Eng. Technol., 1(1) (2016) 116–119. doi: 10.46792/fuoyejet.v1i1.24.

[7] E. Nnochiri and O. Aderinlewo, Geotechnical Properties of Lateritic Soil Stabilized With the Ashes of Oil Palm Fronds, Stavební Obz. - Civ. Eng. J., 25(4) (2016). doi: 10.14311/cej.2016.04.0022.

[8] E. S. Nnochiri and O. M. Ogundipe, Geotechnical Properties of Lateritic Soil Stabilized with Ground-Nut Husk Ash, Civ. Eng. J., 2(11) (2016) 568–575. doi: 10.28991/cej-2016-00000059.

[9] M. Alhassan and A. Mustapha, Effect of rice husk ash on cement stabilized laterite, Leonardo Electron. J. …, 11 (2007) 47–58. [Online]. Available: http://lejpt.academicdirect.org/A11/047_058.pdf?origin=publication_detail.

[10] B. D. Oluyemi and Ayibiowu, Effect Of Cement on Asphalt - Emulsion Stabilised Lateritic Soils, Int. J. Sci. Res. Innov. Technol., 3(5) (2016) 159–172.

[11] G. M. Ayininuola and S. P. Abidoye, Soil Stabilization Using Bitumen Emulsion and Cement Combination as Additive, J. Earth Sci. Eng., 8 (2018) 66–74. doi: 10.17265/2159-581X/2018.02.000.

[12] B. S. 1377, Soils for civil engineering purposes, (1990).

[13] H. M. Jafer, W. Atherton, F. Ruddock, and E. Loffil, Assessing the Potential of a Waste Material for Cement Replacement and the Effect of Its Fineness in Soft Soil Stabilisation, 9(8) (2015) 915–921.

[14] E. C. Igibah, L. O. Agashua, and A. A. Sadiq, Influence of hydrated lime and bitumen on different lateritic soil samples : case study of Sheda-Abuja, Nigeria, 9(1) (2020) 218–222.

[15] Agashua, O. Lucia, and S. Ogbiye, Influence of Cement, Bitumen, and Lime on Some Lateritic Soil Samples as Pavement Material, March, (2018). doi: 10.1088/1757-899X/413/1/012012.

[16] B. S. 1924, Stabilized materials for civil engineering purposes, London, (1990).

[17] A.S.T.M. C39/C 39M, Standard Test Method for Compressive Strength of Cylindrical Concrete Specimens, American Society for Testing and Materials, West Conshohocken, (2009).

[18] Portland Cement Association, Soil-Cement Laboratory Handbook, Skokie, Illinois, (1992). [Online]. Available: http://secement.org/wpcontent/uploads/2017/04/EB052.07s.pdf.

[19] O. E. Oluwatuyi et al., Case Studies in Construction Materials Ameliorating effect of milled eggshell on cement stabilized lateritic soil for highway construction, Case Stud. Constr. Mater., (2018) e00191. doi: 10.1016/j.cscm.2018.e00191.

[20] M. Joel and I. O. Agbede, Mechanical-Cement Stabilization of Laterite for Use as Flexible Pavement Material, J. Mater. Civ. Eng., February, (2011) 146–152. doi: 10.1061/_ASCE_MT.1943-5533.0000148.

[21] Kenya. Ministry of Roads and Public Works, Road Design Manual: Part III Materials & New Pavement Design, (1987) 1–263.

[22] O. W. Road and O. Development, A Guide to The Structural Design of Bitumen- Surfaced Roads in tropical and Sub-Tropical, London, (1993). [Online]. Available: http://insitu-asphalt.com/wp-content/uploads/2009/07/A-Guide-To-The-Structural-Design-Of-Bitumen-Surfaced-Roads-In-Tropical-And-Sub-Tropical-Countries-ORN-31-TRL.pdf.

[23] H. F. Winterkorn and E. C. Chandrasekharan, Laterite Soils and their Stabilization, Highw. Res. Board Bull., 4(2) (1951) 10–29.

[24] A.S.T.M. C 618-05, Standard Specification for Coal Fly Ash and Raw or Calcined Natural Pozzolan for Use in Concrete, West Conshohocken, (2005). [Online]. Available: ftp://ftp.ecn.purdue.edu/olek/PTanikela/To Prof. Olek/ASTM standards/standard specification for coal fly ash and raw or calcined natural pozzolan for use in concrete C 618.pdf.

[25] S. Roy and S. K. Bhalla, Role of Geotechnical Properties of Soil on Civil Engineering Structures, March, (2019). doi: 10.5923/j.re.20170704.03.

[26] A. A. Nur, A. Safuan A., N. Md Noor, and H. Yaacob, Effect of Cement Stabilized Kaolin Subgrade on Strength Properties, J. Appl. Sci., 14(8) (2014) 842–845. doi: 10.3923/jas.2014.842.845.

[27] O. S. B. Al-Amoudi, Characterization and Chemical Stabilization of Al-Qurayyah Sabkha Soil, J. Mater. Civ. Eng., 14(6) (2002) 478–484. doi: 10.1061/(asce)0899-1561(2002)14:6(478).

[28] S. Jaritngam, O. Somchainuek, and P. Taneerananon, Feasibility of laterite-cement mixture as pavement base course aggregate, Iran. J. Sci. Technol. - Trans. Civ. Eng., 38(C1+) (2014) 275–284..

[29] W. W. Bandara, W. K. Mampearachchi, and K. H. S. M. Sampath, Cement Stabilized Soil as a Road Base Material for use in Sri Lankan Roads, (2017) 1–9.

[30] Afolayan., Evaluation of the Effect of Lime and Cement on the Engineering Properties of Selected Soil in a University in Southwestern Nigeria ., J. Adv. Eng. Technol., 5(4) (2017) 1–6.